Short Paper

# Query Game 2.0: Improvement of a Web-Based Query Game for Cavite State University – Main Campus

Mark Philip M. Sy
Department of Information Technology, Cavite State University
markphilipsy@cvsu.edu.ph
(corresponding author)

Christian James M. Historillo
Department of Information Technology, Cavite State University

Allen Cris T. Conde
Department of Information Technology, Cavite State University

Ma. Yvonne Czarina R. Costelo
Department of Information Technology, Cavite State University

## Abstract

*Purpose* – The study aimed to improve the previous study covering a web-based query game for Cavite State University. The study created a new mechanic and gameplay for the students to learn Structured Query Language (SQL). The enhancements also focused on the interactions of one or more students playing the game.

*Method* – The researchers used iterative development process methodology in the development of the study. The system was assessed and evaluated using different testing methods: unit, integration, and system testing. After passing the tests, 90 students of Information Technology and Computer Science program along with 10 IT experts evaluated the system.

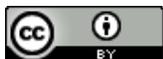

*Results* – The respondents were classified into technical and non-technical respondents and the study garnered an evaluation score of 4.69 and 4.70 respectively. The overall interpretation of the results of the evaluation is Excellent.

*Conclusion* – The study created a system where the user can read and watch lectures and experience the tutorial. Instructors and students may communicate in the system, hence, promoting better relations and healthy competition between students.

*Recommendations* – Based on the conclusions of the study, the system can be used as a supplementary tool in teaching courses with Database Management. To enhance the analysis of query construction of the students, it is recommended to add a module that can analyze the pattern of creating queries and answering questions for every user.

## INTRODUCTION

Learning is a continuous process for everyone. Some people learn better through the use of games. Games can bring different benefits depending on how the users will play and use it. Games are also developed to serve its intended aim. There are games intended merely just for fun and entertainment. While there are also games intended to hasten defending and/or survival skills, critical thinking skills, problem solving skills and so on. These games are often called as educational games. Educational games, from the word itself, are games that have lessons where the player will become more educated about a specific topic.

Today, the trend of developing educational games is high. As mentioned by Burguillo (2010), there is some experimental evidence that computer games increase motivation and can be an effective way to enhance learning. Game developers are now considering not just the fun and entertainment factor that the game will bring to the players but also, the education and knowledge that they can share and learn through the game. But, although educational games are beneficial, the use of these as a new way of learning in school, have never get brought to table and chaired. The web-based query game is a game which lets the player compete with other players with the queries they have learned in database query lessons. The game fetches random question from its database that the players can answer. The players must provide query, particularly MySQL syntax to answer the given question. The study aims to provide an improvement for the web-based query game that will make the system more useful and entertaining to use for students. This will also help the students learn and understand SQL command related matter in a less difficult environment.

## Objectives of the Study

The general aim of this research study was to enhance the Web-based Query Game for Cavite State University – Main Campus.

Specifically, it aimed to:
1. identify the difficulties faced by the students in learning SQL statements, through interviews and survey;
2. analyze the gathered problems through the use of the fishbone diagram;
3. develop a practice mode inside the game module that can guide the students build their own SQL statements;
4. create an information module that can help the students understand the subject matter;
5. develop another game category that will help the students play in a new game mode; and
6. evaluate the system using different testing such as unit, integration, and system testing.

## Theoretical Framework of the Study

Through the study, the students can acquire knowledge about SQL statements in a fun and entertaining way. The students can study a lesson even if their instructors have not yet taught the lesson. They can also assess their understanding by playing the game. The system is a web-based system that utilizes a computer network to facilitate the connection between the server and the user. Figure 1 shows the flow of connectivity of the user to the modules of the system. The system is composed of five (5) modules; account, profile, information, game, and ranking module.

In the profile module, the players can view their own account details on the profile page. Players can also search profiles of other players while logged in inside the game. The information module has two (2) modes, the lecture and tutorial mode. Lecture mode contains text, images, and embedded videos about a specific topic. On the other hand, tutorial mode allows players to see an actual demonstration and simulation of a query syntax execution. The game module is divided into the solo mode and multiplayer mode. In the solo player mode, a player can select between the casual and the custom mode. Casual mode randomly selects 10 questions from the question bank, while the custom mode can be selected to set the number of questions and the level of difficulty. Some questions provide guides such as image and tables in order to help the players form the correct answer. The multiplayer mode allows the player to interact with other active players inside the game. There are two (2) types of room, the Casual and Competition. A minimum of two (2) players are required to start a multiplayer game. Every room can select between the two (2) types of game mode, the single elimination or double elimination. Player-created rooms are casual rooms while the administrator-created rooms are the competition rooms. In single elimination, if the player loses the round, they will be immediately eliminated from the game. While on the double elimination, the player will be placed into the loser's bracket after losing the round. Users can also spectate a game via the spectate option. Players are allowed to access spectator mode if the room allows this option. Chat functions were provided to have communication among the players and administrators. During the game, there will be a time limit to answer the question per round. The goal in the multiplayer mode is to provide the correct answer in the fastest possible time. A player loses once the opponent correctly answers first or when they reach the end of the time limit. During the practice mode, the user will be given their own SQL table where pre-defined SQL commands can be used.

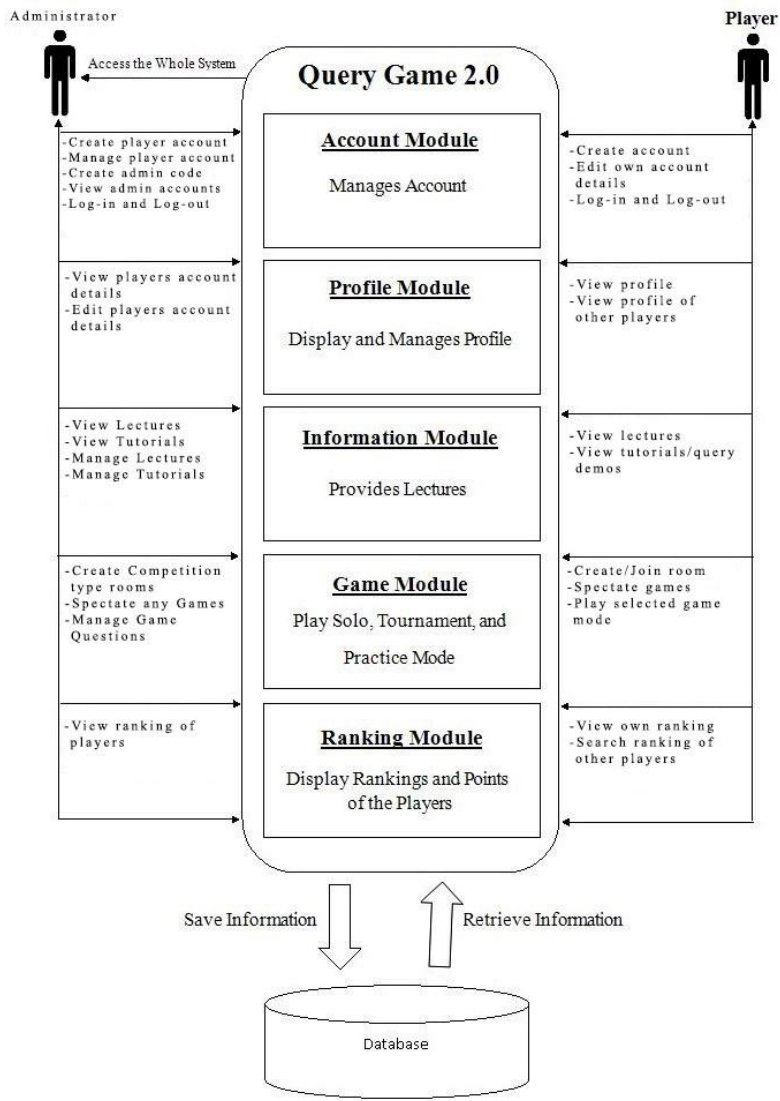

*Figure 1.* Theoretical Framework of Query Game 2.0

# LITERATURE REVIEW

Many studies have been conducted that are focused on the impacts of using games in learning. These studies show different views on the subject matter. In today's educational landscape, entertainment is an additional aspect that may impact learning.

## *Game-based learning*

There have been studies conducted that focuses on the involvement of games in learning. These studies were used as basis in pursuing the main concept of this paper to involve game-based strategy in learning. Even if playing games is often considered a frivolous pastime, gaming environments may actually cultivate a persistent, optimistic motivational style (Granic, Lobel & Engels, 2013). As supported in the study of Rugeli and Lapina (2019), the learning environment can help facilitate learning. Their study about game-based learning used in programming saw that the main advantage of such environments is that they foster their users to learn and keep progressing, making programming fun. Through gaming, users work to achieve meaningful goals, persevere in the face of failures, and celebrate the moments of triumph after successfully carrying out challenging tasks. This fundamental benefit of motivating users through gaming can be harnessed to learn subjects taught in schools.

Through game-based learning that involves collaboration, users can greatly benefit. Devitt (2013) found out that students' interest and enjoyment in playing the math video game increased when they played with another student. In the research, students also view errors and hardships as part of the learning process—rather than manifestations of their lack of ability. This aspect of gaming with communication implemented in learning may reap similar benefits. Shin et al. (2012) conducted two studies to examine the effects of technology-based against a paper-based game in elementary mathematics. They concluded that it was beneficial to students using the technology-based game on all the ability levels of the students when learning arithmetic skills. However, there was insufficient evidence to suggest that technology positively impacted teaching and learning. Through the study of Vandercruysse, Vandewaetere and Clarebout (2012), they have taken an optimistic stance towards the potentials of games in education. The stance is also supported by the research of Shabalina et al. (2017). In their study, they generally think that using game-based learning can be effective for teaching programming.

In the study of Granic et al. (2013), they found out that academic and professional institutions agree that web-based learning environments can give pedagogical advantages. Web-based education tools can increase communication between students and teachers, through discussion boards, chats, and e-mails. The researchers saw that the increase in student motivation and class participation through the inclusion of these elements to a course is possible.

## Related Studies

### Web-Based Query Game Development

The study entitled "Web Based Query Game Development" was developed to help the professors, to teach, and students, to learn, database queries in an easy and enjoyable manner using Randomized Algorithm (Oliveros & Rivero, 2016). This study was the basis for this research and has been a big boost as a predecessor to this endeavor.

### Learning Prosody in a Video Game-Based Learning Approach

This study has taken the video game approach and implemented it for the students to learn prosody. Aguilar (2019) concluded that it was suitable for students due to its particular characteristics. Aguilar stated that "it enhances the learning of soft skills while multimediality (sound, image, text) offers the most appropriate solution for practicing prosody". Audio files can be incorporated to capture the differences between pronunciations and to record the utterances of the players.

### A Research on Applying Game-Based Learning to Enhance the Participation of Student

This study sought to find ways to enhance the participation of students in classes. Lai et al. (2012) stated that even with the use of online platforms (engaging in learning activities online, discussion with other students, and cooperative learning), elevating the motivation of the students using that method alone is difficult. This study devised a plan to add games in studying operating system. This study concluded that "adding game elements in learning operating system courses can increase the level of attraction to the subject". Alejandro (2017) concluded in his research that teachers need to continuously innovate classroom activities; this is to increase the level of participation of the students.

## METHODOLOGY

The hardware used in the study has the following components: 4GB DDR3 random access memory (RAM), Intel (R) Celeron ® processor. The applications used were: Windows 10 Pro 64 Bit (Build 14393) as operating system, PHP and JavaScript as the scripting language, Sublime text editor as the code editor, CSS and Bootstrap in designing the website, MySQL as the database, Adobe Photoshop CS6 as enhancing the design interface and Microsoft Office Word 2016 for documentation.

### Iterative Methodology

The iterative model was chosen because of its simple implementation, which then progressively adds features and functions until the final build. The phases of the Iterative methodology are presented in Figure 2.

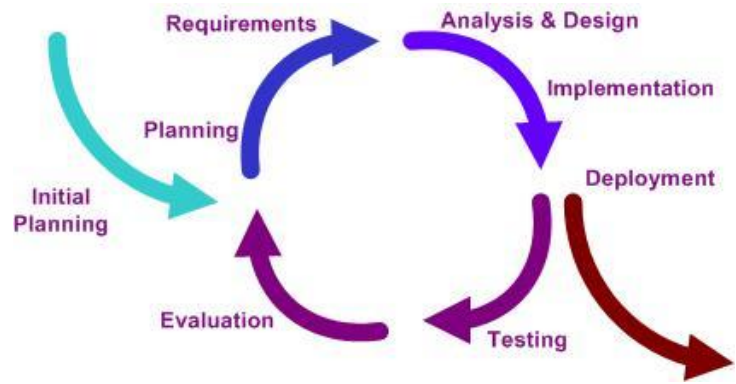

*Figure 2*. Iterative Methodology Phases (Gharai, 2018)

*Planning & Requirements*

In this first phase of the Iterative Methodology, the researches went through the initial planning stage to establish the goals of the study, the software and hardware requirements, and to prepare for the next phases of the cycle. The researchers gathered information by having an interview with the previous study's adviser in order to identify the problems for the improvement of the web-based query game. The researchers also gathered information on the Internet as well as in the library of Ladislao N. Diwa Memorial Library in Cavite State University – Main Campus, Emilio Aguinaldo College, AMA University, and also from College of Engineering and Information Technology Reading Room.

*Analysis & Design*

After the first phase was completed, the researchers analyzed all the problems gathered. In the design phase, the researchers designed the modules that met the needs identified in the analysis phase. The researchers also provided sample user interfaces that can be used in the next phase.

*Implementation*

In this phase, the coding process began. All the gathered information on the previous phases by the researchers including document plans, specifications and software design were coded and implemented into the initial build of the project. At the end of this phase, the researchers had the working system.

*Testing*

The researchers placed the system through a series of tests to identify bugs or issues that remained undetected on the implementation phase. The system was also tested if it meets the planned specification of the study and to confirm if the system is stable.

*Evaluation*

After the system was proven to be stable by extensive testing, the system was evaluated by the target evaluators. The researchers conducted the evaluation with the participation of the students and IT Experts of CEIT. Upon receiving the feedback of the respondents, iterations were also made to include some of the suggestions and comments.

**RESULTS**

## Profile of the Respondents

Evaluation was conducted with the participation of the students and IT Experts of CEIT. The respondents were 90 IT/CS students and 10 IT Experts. The selection of the respondents was done through the use of purposive sampling. The respondents were students who have taken the Database Management System subject and instructors that are knowledgeable in SQL.

## Treatment of the Data

The respondents evaluated the software with the set criteria such as functionality, reliability, usability, efficiency, maintainability, portability, user-friendliness. The 5-point Likert scale was used with the following interpretations used: 5-Excellent, 4-Very Good, 3-Good, 2-Fair, 1-Poor. Mean and Standard Deviation were used in interpreting the results of evaluation.

## Software Evaluation

After the completion of the system, the researchers proceeded with the evaluation. The result of the evaluation of the respondents was then tabulated. Table 1 shows the summary of results and interpretation for the technical evaluation that obtained a grand mean of 4.69 and interpreted to be "Excellent".

Table 1. Summary of evaluation for technical evaluation

| Indicators | Average mean | Interpretation |
| --- | --- | --- |
| Functionality | 4.85 | Excellent |
| Reliability | 4.73 | Excellent |
| Usability | 4.57 | Excellent |
| Efficiency | 4.7 | Excellent |
| Maintainability | 4.47 | Excellent |
| Portability | 4.67 | Excellent |
| User-Friendliness | 4.87 | Excellent |
| **Grand Mean** | 4.69 | Excellent |

Table 2 shows the summary of results and interpretation for the non-technical evaluation that obtained a grand mean of 4.69 and interpreted to be "Excellent".

Table 2. Summary of evaluation for non-technical evaluation

| Indicators | Average Mean | Interpretation |
|---|---|---|
| Functionality | 4.71 | Excellent |
| Reliability | 4.75 | Excellent |
| Usability | 4.65 | Excellent |
| User-Friendliness | 4.71 | Excellent |
| **Grand Mean** | 4.70 | Excellent |

As the mean from the technical and non-technical evaluation showed the interpretation Excellent, it implies the system has the potential to change the current way how database managements system, specifically SQL, is taught in the University. This study can likely help students learn more about SQL. This system can also be used as a supplementary tool for teaching SQL.

## DISCUSSION

Query Game 2.0 was developed to improve the Web-Based Query Game of Cavite State University. The study aimed to help students learn SQL queries in a more entertaining way and to create an interactive gaming environment with a multiplayer game mode where players can play together and be matched up in real time.

Figure 3 displays the main page of the game. It is the landing page of the game where users may log in or register an account. The administrator can also use this page to navigate to the administration console.

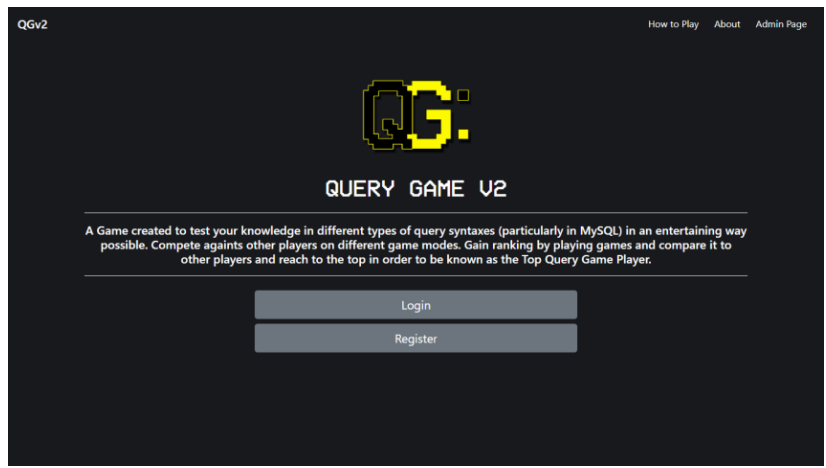

*Figure 3.* Screenshot of the Query Game 2.0 main page

Figure 4 shows the lecture page. Players can see the available lectures that they can use to study SQL. The lectures may contain text, images, audio, and/or videos.

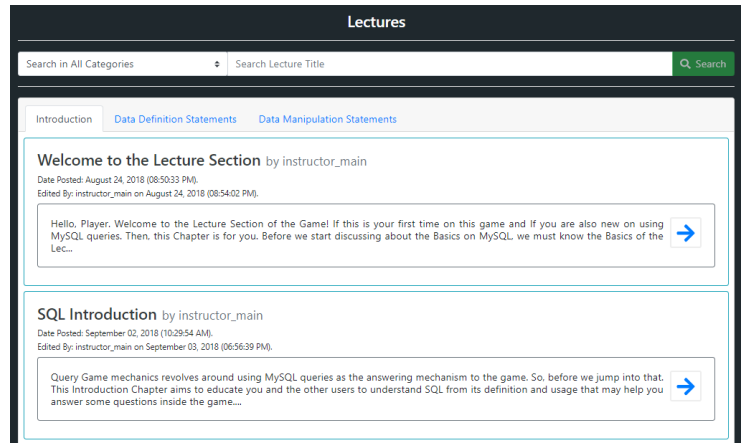

*Figure 4.* Screenshot of the lectures page

Figure 5 shows the profile page of a player. The page contains details about the player such personal information, game mode rankings and the game records of the player. The player can search for other players by using the search function inside the profile page.

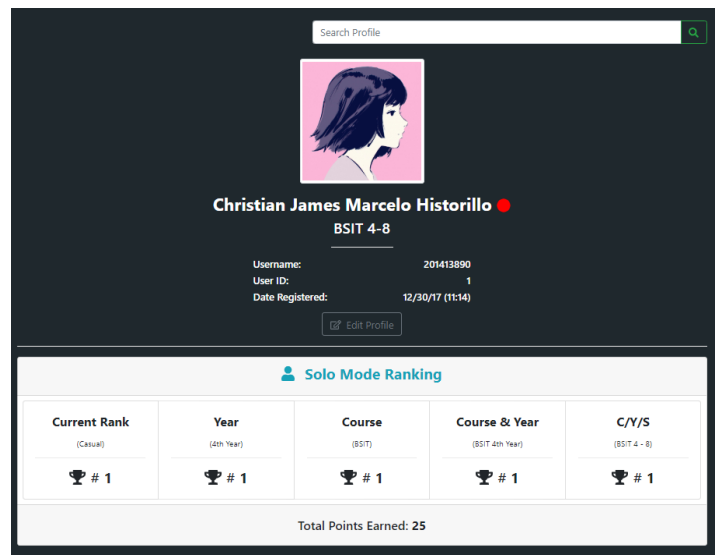

*Figure 5.* Screenshot of the profile page

Figure 6 shows the interface inside the solo mode game. To finish the game, the player can either answer all the questions or leave the game. The questions are randomly generated. Some questions provide additional guides such as tables and image in order to help the user give the correct answer. The player may also skip a question. Before submitting the all of the answers, skipped questions may still be answered while changes to other questions may also be done. After the submission of the answers, a simple evaluation of the score is displayed. The system can display the types of SQL commands where the user excels and where the user is finding a hard time.

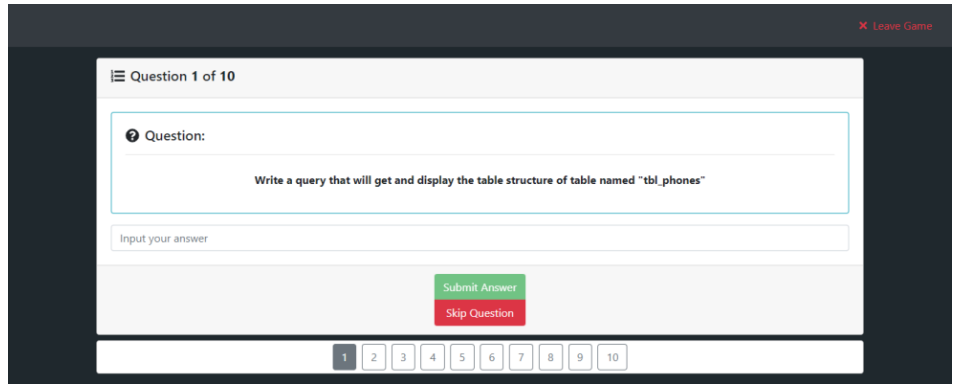

*Figure 6.* Screenshot of the solo mode game interface

Figure 7 shows the creation of rooms for multiplayer game mode. The player must fill the required details in order to create a multiplayer room. The user can also join an existing game room. Available game rooms are listed in the lobby. For student-created rooms, they are casual game rooms, while instructor-created rooms are competition rooms.

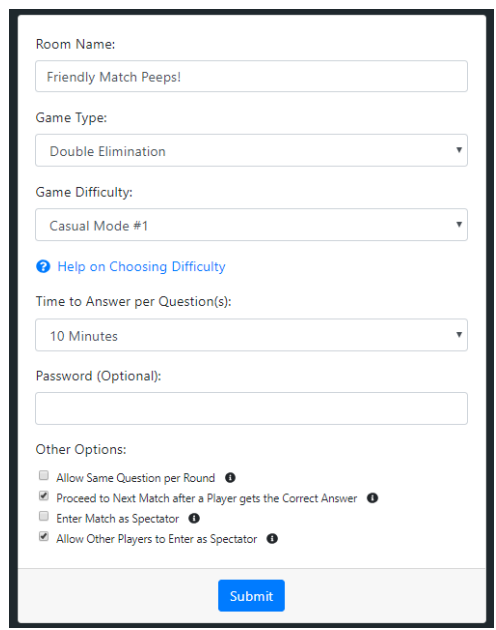

*Figure 7.* Screenshot of how to create multiplayer room

Figure 8 illustrates the page of the multiplayer room. After the player has created or joined a room, they can see the room details (name, type, room settings, etc.). Every player can also chat and see other players. Only the room creator can start the game. The system will automatically pick the match-up based on the provided room details (ex. Game type). Other players may also be included in the room as spectators. Spectators are not picked by the system for bracket matching.

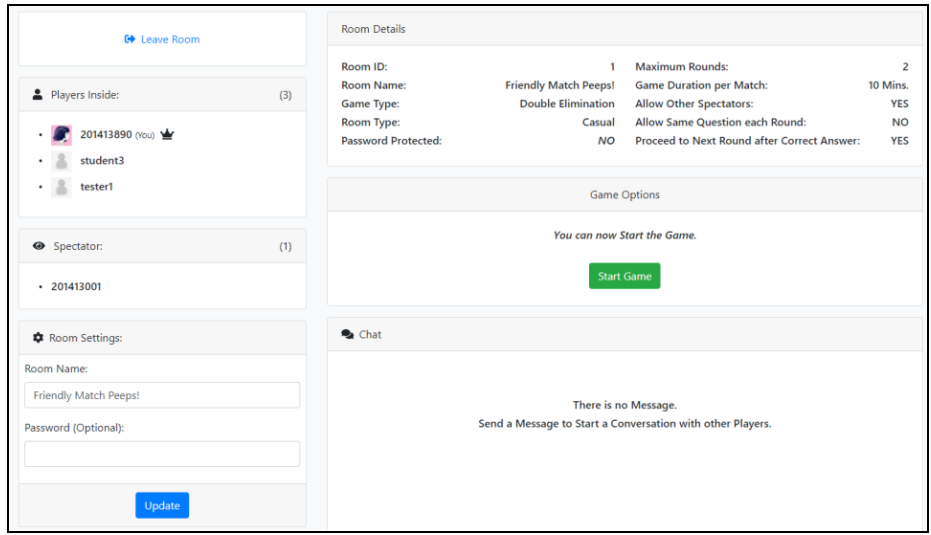

*Figure 8.* Screenshot of the multiplayer room page

Figure 9 shows the screenshot of the question panel. The administrators can manage the questions used in the solo player mode. The administrators can view, add, edit and delete questions.

*Figure 9.* Question Panel

To facilitate the execution of the queries by the users, submitted queries are executed in the host server. The database of the server contains tables of the system along with tables for the game. The tables for the game are built-in along with the questions, specifically for the multiplayer game mode. For the practice mode, players are allowed to execute their queries on their own tables. Each account will have a pre-built table named after the user (i.e. 201910001_table). Each user can manipulate the contents and the structure of their table. As shown in Figure 10, submitted queries will be handled by a query processor, a script that enforces rules on queries.

A query processor for the practice mode will limit the queries to ALTER, TRUNCATE, and basic DML commands (SELECT, INSERT, UPDATE, DELETE). Queries with DROP or CREATE keywords, along with other DCL commands are not executed. A query will be parsed for keywords that may cause harm to the existing database structure and will not be executed. For the solo player mode, queries for the questions are already stored. Once the user enters their query, it will be compared to the stored string. Most of the questions in the solo player mode focus on basic queries. In the multiplayer game mode, there are stages or rounds.

Questions that ask for DML commands (except SELECT) rely on simply accepting queries in the query processor then checking if the answer matches the stored answers. For other questions especially SELECT statements, it uses a shadow table. Knowing that a question can have multiple correct queries, these questions are processed differently and not directly compared to the stored answers. Queries for complex questions are run with another shadow table. The shadow table contains the same structure as the given problem, but with a different table name and set of data. After the execution of the query on the shadow table, the result is converted into a string and the generated string is compared to the stored answers. Through this way, the player cannot create a direct query that will only get values on the given table (creating a SELECT statement that fetches records asked in the question). This type of blind test can ensure that the player created a query that is compliant with the question. Every player can submit an answer as long as the correct answer is not yet provided by their opponent. To facilitate the real-time conversation in the chat and the answers in the question, WebSockets was used.

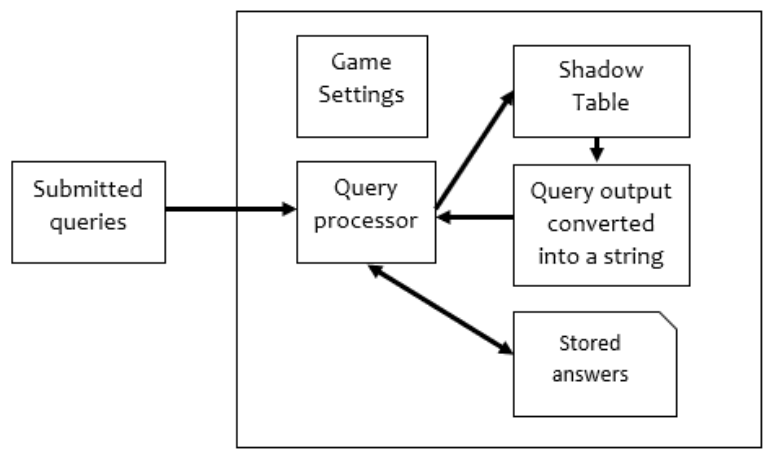

*Figure 10*. Process flow in the game modes

# CONCLUSIONS AND RECOMMENDATIONS

The system solved the problems identified in the planning and requirements phase. The system can be played with one or more students. The administrator (professor) can create accounts for other administrators (professor), give verification code for the newly created accounts of students and also create a session game. The interactions of students and administrators (professor) were enhanced. The game learning information was properly placed and is more understandable. Newly added modes can be played within the game. More and new questions are being organized prior to their difficulty and the ranking of students can be seen within the system. Different queries, particularly on SELECT type of queries that has the same answer can be detected by the system to check if the user answer is correct or not. On DML queries, the administrator could place different answer combinations to provide more answer checking possibilities to the system. The study has shown that with the use of game-based learning, it can greatly benefit users.

Ranking module provide more details per game mode that helps the players identify their own ranking and the other player ranking on different game mode areas. After a series of tests conducted by the researchers, the system met its objective and now ready for deployment in order for the CEIT to make their tasks of teaching SQL easier. This study has shown that technological advancements can be used to enhance the existing ways of learning SQL especially in a University setup. It is highly recommended that the system be put in place in the curriculum in the course that tackles Database Management System. Even though the system is considered Excellent, based on the evaluation, to further improve it, it can include a module that analyzes the pattern of creating queries and answering questions for every user. In this way, a deeper analysis of how a user understands, formulates and learn a query can be studied.

# IMPLICATIONS

Based on the findings of the study, the developed system is a good supplementary tool for the curriculum that includes Database Management System, specifically those that cover the construction of SQL queries. However, the study only covered the development and evaluation of the system. To further the findings of the research, a study that measures the efficacy of the system versus the traditional face-to-face learning may be conducted. This study has already established a tool that may be used to improve the experience of the students learning queries.

# ACKNOWLEDGEMENT

The researchers would like to thank Cavite State University for helping out and allowing this study to be conducted. This undertaking will not be possible without the help of the College of Engineering and Information Technology faculty and staff, particularly to the Department of Information Technology. To the students that served as respondents and for giving their honest opinions and feedback towards the study, to the CvSU Research

Center for helping the researchers through the support they have provided, and to the parents, siblings, and friends of the researchers for being there and being an inspiration. Above all, to the Almighty God for helping the researchers in every means possible. This endeavor is dedicated to all of them.